\documentclass[aps,prl,twocolumn,amsmath,amssymb,superscriptaddress,nobibnotes,showpacs,floatfix]{revtex4-1}
\usepackage{graphicx}
\pdfoutput=1
\usepackage{ulem}
\usepackage{amsmath}
\usepackage{amssymb}
\usepackage{amsthm}
\usepackage{tabularx}
\usepackage{multirow}

\usepackage[usenames, dvipsnames]{color} 

\usepackage{hyperref}
\hypersetup{
  colorlinks = true,    
  allcolors = blue
}

\def\vR{{\bf R}}

\begin{document}

\title{Modelling Magnetic Multipolar Phases in Density Functional Theory}

\author{Dario Fiore Mosca}
\affiliation{University of Vienna, Faculty of Physics and Center for Computational Materials Science, Vienna, Austria}

\author{Leonid V. Pourovskii}
\affiliation{Centre de Physique Th\'eorique, Ecole Polytechnique, CNRS, Institut Polytechnique de Paris, 91128 Palaiseau Cedex, France}
\affiliation{Coll\`ege de France, 11 place Marcelin Berthelot, 75005 Paris, France}

\author{Cesare Franchini}
\affiliation{University of Vienna, Faculty of Physics and Center for Computational Materials Science, Vienna, Austria}
\affiliation{Department of Physics and Astronomy "Augusto Righi", Alma Mater Studiorum - Universit\`a di Bologna, Bologna, 40127 Italy}

\date[Dated: ]{\today}

%

\begin{abstract}
Multipolar magnetic phases in correlated insulators  represent a great challenge for Density Functional Theory (DFT) due to the coexistence of intermingled interactions, typically spin-orbit coupling, crystal field  and complex non-collinear and high-rank inter-site exchange, creating a complected configurational space with multiple minima.
Though the +U correction to DFT  allows, in principle, the modelling of such magnetic ground states, its results strongly depend on the initially symmetry breaking, constraining the nature of order parameter in the converged DFT+U solution. 
As a rule, DFT+U calculations starting from a set of initial on-site magnetic moments  result in a conventional dipolar order. 
A more sophisticated approach is clearly needed in the case of magnetic multipolar ordering, which is revealed by a null integral of the magnetization density over spheres centered on magnetic atoms, but with non-zero local contributions.
Here we show how such phases can be efficiently captured using an educated constrained initialisation of the onsite density matrix, which is
derived from the   multipolar-ordered ground state of an ab initio effective Hamiltonian.
Various properties of such exotic ground states, like their one-electron spectra, become therefore accessible by all-electron DFT+U methods.
We assess the reliability of this procedure on the Ferro-Octupolar ground state recently predicted in Ba$_2$MOsO$_6$ (M = Ca, Mg, Zn) [Phys. Rev. Lett.\textbf{127}, 237201 (2021)]. 
\end{abstract}

\maketitle

Transition Metal Oxides (TMOs) with strong Spin Orbit Coupling effect (SOC) have attracted great attention due to the realization of unconventional magnetic phases, ranging from canted antiferromagnetic (AFM) orders observed in Ba$_{2}$NaOsO$_{6}$ ~\cite{Lu2017, Daje,PhysRevB.82.174440} and Sr$_{2}$IrO$_{4}$ ~\cite{PhysRevB.92.054428} to high-rank magnetic multipoles ~\cite{RevModPhys.81.807, RevModPhys.83.1301}. Remarkable examples of multipolar orderings have been reported for URu$_{2}$Si$_{2}$ and NpO$_{2}$, where the onset of hidden ordered phase transitions have been 
connected with possible transitions towards a multipolar magnetic phase ~\cite{RevModPhys.83.1301, Pourovskiie2025317118}. 
More recently, 5d-based TMOs have attracted considerable interest due to the interplay between an unexpectedly high electronic correlation and SOC ~\cite{PhysRevLett.102.017205}, with several reports providing evidence on the possible formation of multipolar ground states ~\cite{Hosoi, PhysRevB.101.155118, JPSJ.90.062001}.
The majority of these works are based on microscopic low-energy effective Hamiltonians, solved by a variety of many-body methods.
In fact, the search of multipolar magnetism by means of DFT electronic structure schemes with the +U correction inevitably faces the problem 
of being trapped in local minima corresponding to  conventional dipolar solutions. In a pioneering work S.-T. Pi and coworkers~\cite{PhysRevLett.112.077203} addressed this problem
by calculating exchange interactions through flipping of the expansion coefficients of the onsite matrix expanded in terms of multipolar tensor components. The change in band energies reflects the energy cost of the corresponding flipping that can be afterwards mapped onto the exchange constant via Andersen force theorem (FT) ~\cite{Andersen}. However, this method becomes computationally intensive for multipolar "hidden" order systems, where the space of possible order parameters is large.

An alternative approach~\cite{PhysRevB.94.115117} is based on a FT  formulated for the symmetry-unbroken paramagnetic electronic structure. The latter is obtained within the DFT+DMFT (dynamical mean-field theory~\cite{RevModPhys.68.13,lichtenstein_dft_dmft}) framework using the quasiatomic Hubbard-I (HI) approximation. Even for complex "hidden-order" systems, the full magnetic Hamiltonian can be derived using this FT-HI method from  post-processing of a single DFT+HI calculation for the paramagnetic state \cite{Pourovskiie2025317118,pourovskii2021}. However, in contrast to DFT+U, the DFT+HI method cannot directly model the electronic structure of  multipolar-ordered phases, since leading inter-site interactions in correlated insulators arising through hybridization of localized electrons (e.~g. superexchange) are neglected in DFT+HI.


In this work, we 
develop a framework for calculating multipolar-ordered phases with the DFT+U method by  initializing those calculations using the  output provided by the FT-HI effective-Hamiltonian   method. 
Within our scheme we first identify competing  multipolar phases from an  {\it ab initio} effective Hamiltonian. 
In this Hamiltonian, the relevant ground-state multiplet (GSM) of low-energy electronic states is represented by a pseudo-spin; the inter-site interactions between various moments of this GSM space are calculated by the FT-HI method \cite{PhysRevB.94.115117} from the paramagnetic DFT+HI electronic structure. 
The effective Hamiltonian is then solved either by mean-field or by more sophisticated many-body techniques to obtain the transition temperatures and order parameters for low-temperature phases. 

Using explicit (Fock state) representations of the GSM many-electron states, as calculated by DFT+HI, one may transform those order parameters into on-site one-electron density matrices (ODM).
Then, the DFT+U+SOC run is initialized with such ODMs corresponding to a chosen multipolar order. The selective initialization of the ODM allows the direct total energy calculations of a specific multipolar magnetic ground state, thus avoiding the risk of falling in a metastable dipolar state.

This 
framework enables the study of materials-specific electronic and magnetic properties of a genuine multipolar state in DFT at the atomic scale without adjustable parameters, and allows for a DFT-based analysis of the response of the multipolar ground state to external stimuli such as local structural distortions or doping, which are difficult to treat at DFT+HI level.  

We employ the proposed computational protocol to study the competition between conventional dipolar and magnetic multipolar order in the cubic 5d$^2$ double perovskites (DPs) Ba$_2$MOsO$_6$ (M = Ca, Mg, Zn) (BCOO, BMOO and BZOO from now on). 
With a $t_{2g}^2$ configuration exhibiting a S=1 spin state and an effective orbital moment $l$=1, the low energy physics of these spin-orbit coupled systems can be represented  by a total effective momentum  (pseudo-spin) J$_{eff}$=2, analogous to a single $d$-electron $l=2$ level. As a consequence, in a cubic symmetry, the J$_{eff}$=2 level splits due to the remnant crystal field (RCF) into a lower $E_g$ doublet and a higher-in-energy $T_{2g}$ triplet. Since the non-Kramers $E_g$ doublet is isomorphic to a $e_g$ doublet, it carries no dipole moment,
thus representing an ideal playground for the realization of high-rank multipole orders.

For these reasons BMOO DPs have recently been in the spotlight, but with conflicting experimental data. On one side muon spin resonance  and thermodynamic anomalies show a clear phase transition below T$^*$~$\approx$~30-50~K, with broken time reversal symmetry and with large antiferromagnetic Curie-Weiss constant ($\Theta_{CW} \approx 130$ K) ~\cite{PhysRevB.94.134429, Thompson_2014},  apparently consistent with a weak Néel spin ordering. 
On the other side no magnetic Bragg peaks were observed in neutron diffraction experiments, establishing an upper limit for the Os dipolar magnetic moment of  $\approx 0.1 \mu_{B}$ ~\cite{PhysRevLett.124.087206}. Furthermore, possible quadrupolar orderings are also ruled out by the absence of tetragonal distortion as verified by x-ray diffraction measurements, up to $\approx$ 0.1 \% of the volume, ~\cite{PhysRevLett.124.087206}.

To shed some light on this complex scenario a few 
theoretical analyses have been reported providing robust evidence for the formation of a higher-rank order of the octupolar type~\cite{PhysRevB.104.174431,
PhysRevB.102.064407, pourovskii2021}.
We have recently evaluated the effective many-body Hamiltonian for these compounds using the FT-HI method \cite{pourovskii2021} and obtained a ferro alignment of the $xyz$ octupoles as their ground (ferro-octupolar, FO) state \cite{pourovskii2021}. 
This FO order is enabled by a large RCF (an order of magnitude larger than the intersite exchange interaction) suppressing competing dipolar orders
and mediated by superexchange mechanism through O-p and Ba orbitals.~\cite{pourovskii2021}. 
An anti-ferro order of quadrupoles active within the $e_g$ doublet was identified in Refs.~\cite{pourovskii2021,PhysRevResearch.3.033163} as a competing phase. 
However, neither the electronic structure of the FO state nor its competing phases has been calculated in  Ref.~\cite{pourovskii2021} due to the above mentioned limitations of DFT+HI. 

In this work we make use of the previously obtained FT-HI Hamiltonians and ordered phases to carry out electronic structure calculations for the $d^2$ DP series  using the constrained DFT+U methodology outlined above. We identify signatures of the multipolar order in one-electron  spectrum.  We also show that this DFT+U methodology is able to qualitatively  capture the energetics of multipolar orders; in particular, it  correctly predicts the relative magnitude of ordering energies along the  5d$^2$ DP series.


\section{Method}

We 
carried out DFT+U calculations 
by the Vienna Ab initio Simulation Package (VASP) ~\cite{PhysRevB.54.11169,PhysRevB.47.558} using the generalized gradient approximation of Perdew, Burke and Ernzerhof. We included the onsite Coulomb repulsion at the Os $d$ shells using the
rotationally-invariant
Lichtestein formulation of DFT+U~\cite{PhysRevB.52.R5467}. The on-site Coulomb vertex is specified by the Hubbard U  and Hund's rule coupling J; we employed U = 3.2 eV and J = 0.5 eV , in agreement with previous works~\cite{pourovskii2021}. 
The spin-orbit coupling  was included in DFT and we estimated, via exact diagonalization of the atomic levels in the local Hamiltonian, the resulting SOC strength $\lambda$  to be $\sim 0.3$ eV.
For both BCOO, BMOO and BZOO the experimental lattice structures were used from references ~\cite{Thompson_2014} and ~\cite{PhysRevB.94.134429} respectively; the reciprocal space was sampled with a 6~$\times$~6~$\times$~6 k-mesh and an energy cutoff for the plane wave expansion of 600 eV was applied.


We employ the protocol outlined in the introduction 
to construct an 
appropriate
 ODM for a given multipolar order. 
This  ODM is then used as  starting guess for the constrained-ODM implementation of reference~\cite{C4CP01083C}. 
The starting ODM  is derived from the order parameters of a given multipolar phase. To that end we represent the one-electron ODM corresponding to a given set of the order parameters in the pseudo-J space as follows
\begin{equation}
     \rho_{_{HI}}^{mm',\alpha} = Tr \big[ \rho^{mm'}_{MM'} (J) \   \hat{\rho}_{\alpha}(J) \big] 
\end{equation}
where   $\rho^{mm'}_{MM'} (J) = \langle JM |c^{\dagger}_{m}   c_{m'} |JM' \rangle $ is the $mm'$ matrix element 
of the ODM operator in the  GSM basis $|JM\rangle$, $c^{\dagger}_{m}, c_{m'}$ are the one electron creation and annihilation operators and $\hat{\rho}_{\alpha} (J)$ is the many-electron GSM density  matrix (DM) at the site $\alpha$. The DM $\hat{\rho}_{\alpha} (J)$ is computed from the ordered moments at site $\alpha$ for a given phase,  $\hat{\rho}_{\alpha} (J)=\sum_{KQ}\hat{O}_K^Q(J) \langle \hat{O}_K^Q(J)\rangle_{\alpha}$, where O$^{Q}_{K}$ is the spherical Hermitian  tensor for given J  with rank $K = 1...2J$ and projection $Q$ ~\cite{RevModPhys.81.807} and $\langle \hat{O}_K^Q(J)\rangle_{\alpha}$ is its expectation value at the site $\alpha$ in a given ordered state. For the d$^2$ DP series  we use the FO order parameters $\langle \hat{O}_K^Q(J)\rangle$  that were obtained in Ref.~\cite{pourovskii2021} by solving the corresponding FT-HI effective Hamiltonians for pseudo-spin $J=$2. To evaluate the matrix elements $\langle JM |c^{\dagger}_{m}   c_{m'} |JM' \rangle$ we employ the many-electron  states of the GSM  obtained in DFT+HI calculations~\cite{pourovskii2021} of those compounds; these states are expressed in the $ms$ $d$-electron Fock basis rendering such evaluation straightforward.  

\begin{figure}[!h]
  	\begin{centering}
  		\includegraphics[width=1\columnwidth]{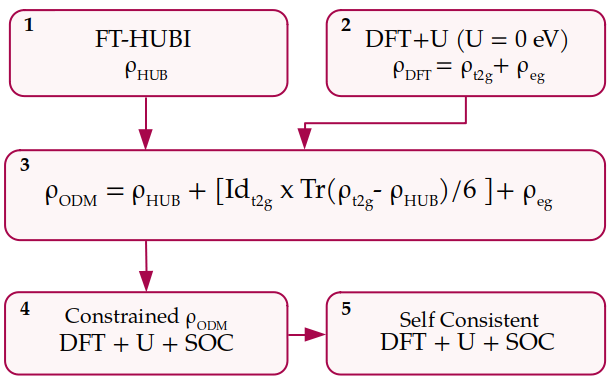} 
  		\par\end{centering}
  	\caption{Scheme of the constrained multipolar ODM protocol: 1) FT-HI optimization and calculation of the FT-HI ODM $\rho_{_{HUB}}$; 2) Preliminary DFT + U (= 0 eV)  calculation for the t$_{2g}$ and e$_{g}$ density; 3) Hybridization correction and initialization of a $\rho_{_{ODM}}$ consistent with  $\rho_{_{HUB}}$; 4) ODM-constrained DFT+U+SOC calculation with the new $\rho_{_{ODM}}$; 5) Full self-consistent DFT+U+SOC run starting from the pre-converged wavefunction generated in (4) (see SM~\cite{supplmat}).}
  	\label{fig:0} 
  \end{figure}

 The many electron DM $\hat{\rho}_{\alpha} (J)$ contains the nominal number of correlated electrons included in the effective Hamiltonian. DFT calculations, on the other hand, typically overestimate this counting because of the strong hybridization between the $d$ orbitals with the O $p$ states. 
 This enhanced electron counting is particularly problematic for osmates double perovskites, where the formal 5d$^2$ occupation is actually $\sim$ 6 electrons in DFT~\cite{Daje,PhysRevB.91.045133}. This unbalanced treatment of the ODM in DFT ($\rho_{_{ODM}}$) is incongruous with the nominal $t_{2g}^2$ configuration adopted in the FT-HI effective Hamiltonian ($\rho_{_{HI}}$) and needs to be corrected in order to have a consistent mapping between the two models.
 In our protocol (see Figure ~\ref{fig:0}) we correct this problem by quantifying the nominal excess charge from a preliminary spin-unpolarized DFT calculation (U=0), were we obtain a reference DFT-ODM ($\rho_{_{DFT}}$) which is a sum of the contribution coming from t$_{2g}$ orbitals ($\rho_{_{t2g}}$) and the e$_g$ ones ($\rho_{_{eg}}$) that include both the nominal d$^2$ electrons as well as additional electronic charge coming from hybridization effects.

As in VASP the $\rho_{_{DFT}}$ is calculated in the global coordinate system one has to be sure this matches the local reference frame of the octahedral environment, to avoid mixing contributions from different orbitals. In our unit cells these reference frames  coincide, and such a rotation is not needed. 
However, another change of basis was employed in order to move from the spherical harmonic basis of the $\rho_{_{HUB}}$ to the cubic one defined in VASP (see Supplementary Materials (SM)~\cite{supplmat}).
At this point the the hybridization contribution can be readily obtained by splitting the correlated $t_{2g}$ part from the e$_g$ one and defining
\begin{equation}
    \rho_{_{ODM}} = \big[ \rho_{_{HI}} + Id_{t2g}  \times Tr(\rho_{_{t2g}} - \rho_{_{HI}})/6 \big] + \rho_{_{eg}} \  ,
\end{equation}
where $Id_{t2g}$ is the identity acting on the t$_{2g}$ subspace only and the division by 6 is due to the spin degeneracy.

With the new ODM that correctly counts the added electrons, we have an educated initial guess on top of which we were able to run a ODM-constrained calculation at DFT+U+SOC level, with the requirement of keeping the total magnetization along the global components as obtained with $Tr(S_{i} \rho_{_{ODM}})$, where $S_{i}$ is the dipolar spin operator. 
As mentioned above, the nature of the FO ground state is such that all these components average to zero and as such have to be initialised, considering that any other initialization would act as local magnetic field on the osmium atom, thus pushing towards a conventional dipolar solution.
After the initial ODM-constrained calculation, we performed a consistency safety check to control that the correct number of electrons is maintained in the ODM, and subsequently performed a full self-consistent calculation to obtain the FO DFT+U+SOC solution, starting from the pre-converged wavefunctions.
A consistency check is also done at the very end of the self-consistent in order to check that the $Tr(\rho_{_{ODM}})$ is compatible with the required initialization.

The interest in the nature of the FO phase concerns the potential  differences with respect to conventional dipolar solutions. To gain a better understanding, we performed a series of DFT+U+SOC calculations with multiple different dipolar configurations from FM to AFM (see SM~\cite{supplmat}), finding the type-I AFM-110 as lowest energetic dipolar one. From now on we will use the AFM-110 solution as measurement of comparison, where AFM-110 means AFM arrangement of dipolar magnetic moments along parallel planes in the [001], with magnetic moments lying along the [110] crystallographic direction.

\section{Results}

We start by discussing the FO solution obtained by DFT using the proposed approach and comparing it with the dipolar AFM-110 state. After that, we present the DFT estimations of the intersite exchange couplings and explore the possibility for tetragonal transition.

\subsection{The Ferro-octupolar phase}

To extract useful information from the DFT output we fit the final DFT-ODM $\rho_{_{ODM}}$ to the basis of 2-electrons average of multipolar moments $\langle O_{m}^{n} (J) \rangle$:
\begin{equation}
    \rho_{_{ODM}} = \sum_{n,m} a_{m}^{n} \ \langle O_{m}^{n} (J) \rangle,
\end{equation}
where $a_{m}^{n}$ are the fitting coefficients. 

\begin{table}[b]
	\begin{center}
		\begin{ruledtabular}
			\renewcommand{\arraystretch}{1.2}
			\begin{tabular}{l | c c c c }
				Compound & Tr($\rho_{_{ODM}}$) &  $O^{0}_{2}$ & $O^{2}_{2}$ & $O^{-2}_{3}$  \\
				\hline
				Ba$_{2}$CaOsO$_{6}$  & 6.04 & 0.15 & 0.25 & 0.51 \\
				
				Ba$_{2}$MgOsO$_{6}$ & 6.04 & 0.14 & 0.17 & 0.56 \\

				Ba$_{2}$ZnOsO$_{6}$  & 6.34 & 0.15 & 0.22  & 0.58 \\

			\end{tabular}
		\end{ruledtabular}
		\caption{\label{Tab:1}  Charge on the Os atoms and values of the saturated multipolar moments for the different compounds in the FO phase, with 1 is the fully saturated moment.}
	\end{center}
\end{table}

  \begin{figure}[!tb]
  	\begin{centering}
  		\includegraphics[width=1\columnwidth]{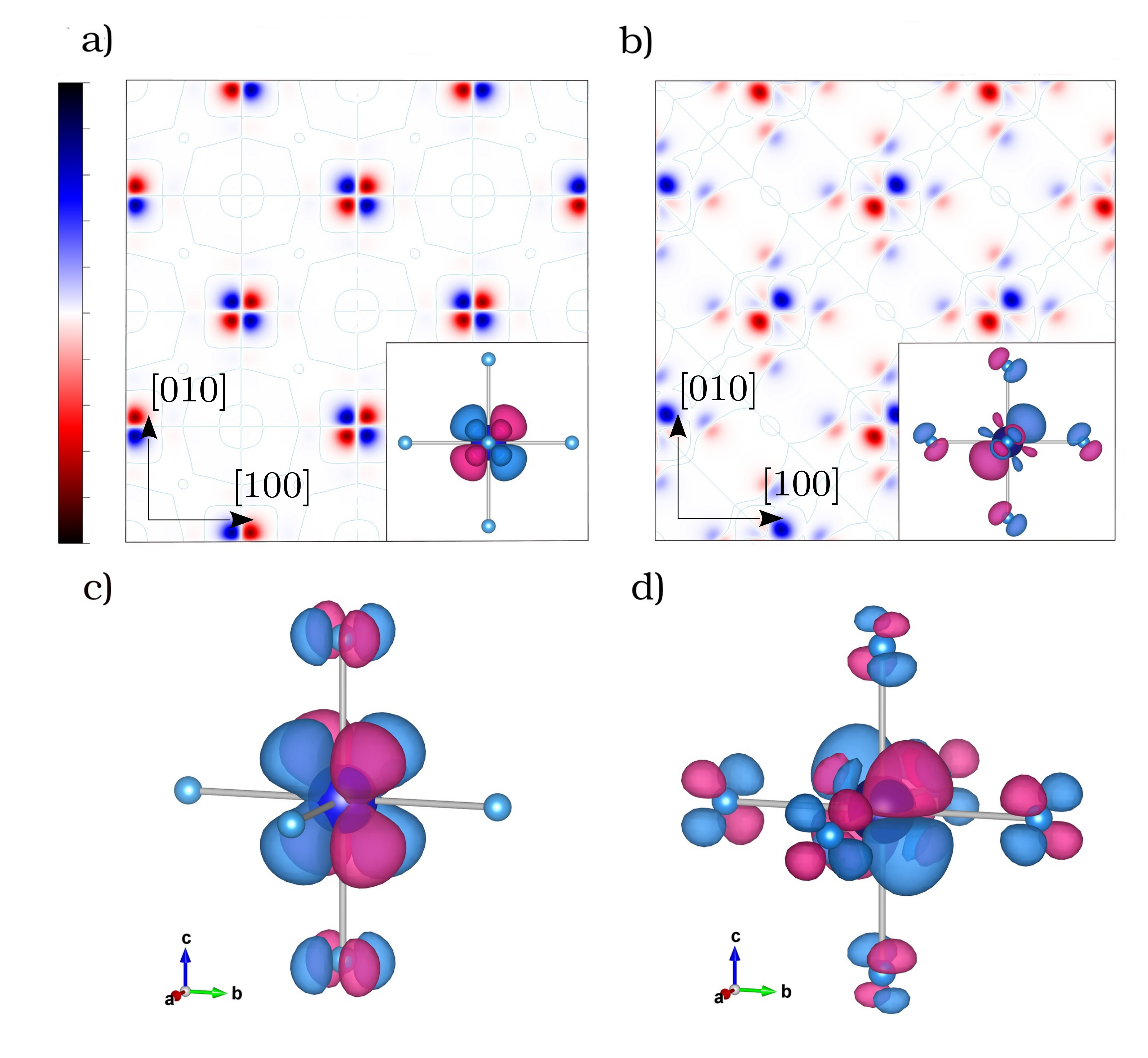} 
  		\par\end{centering}
  	\caption{Color plot of the magnetization density along the z direction on the Os sites as seen from the [001] crystallographic direction for the FO (a) and AFM-110 (b) cases. The 3-dimensional plots are (c) and (d) respectively. The complete plots along the x and y direction are given in the SM~\cite{supplmat}.}
  	\label{fig:1} 
  \end{figure}

We find indeed that the octupolar $O^{-2}_{3}$ operator remains non zero, even if not fully saturated (see Table ~\ref{Tab:1}), providing clear evidence of the capability of DFT to model a genuine multipolar order.
Most of the other magnetic multipoles are almost zero, apart from $O_{2}^{2}$, $O_{2}^{0}$ and hexadecapoles, the latter with values that do not impact on the quality of the fit and are therefore neglected from now on.
The non zero values obtained for quadrupolar $O_{2}^{2}$ and $O_{2}^{0}$ terms are a consequence of switching-off all symmetry as required in VASP-based SOC calculations, and do not imply any tendency of the system to undergo a cubic-to-tetragonal transition (we have verified that the systems prefers to preserve the cubic symmetry, see Sec. "Tetragonal Distortions"). 
We further confirmed the role of symmetry by performing a non-magnetic DFT + SOC calculation from which we extracted the expansion coefficients finding values of $\sim$ 0.30 for $O_{2}^{0}$ and $\sim$ 0.43 for $O_{2}^{2}$, then subtracted from the pure DFT + U + SOC values to partially compensate the above-mentioned effect (see Table \ref{Tab:1} for the re-scaled values).
The final ODM contains also spurious terms coming from the hybridization between Os and O atoms, such that only 90 \% of the total Hilbert space from which $\rho_{_{ODM}}$ is constructed can be correctly mapped in our the tensor fit;  this unbalance can be easily adjusted by subtracting the the non-magnetic DFT+SOC ODM calculation from the final ODM.
A visual representation of the obtained FO state is given in Fig.~\ref{fig:1} in terms of the magnetization density isosurface along the crystallographic $z$ direction, showing a FM order of magnetic octupoles. 

\begin{figure*}[t]
  \begin{center}
  \includegraphics[width=1.0\textwidth,clip=true]{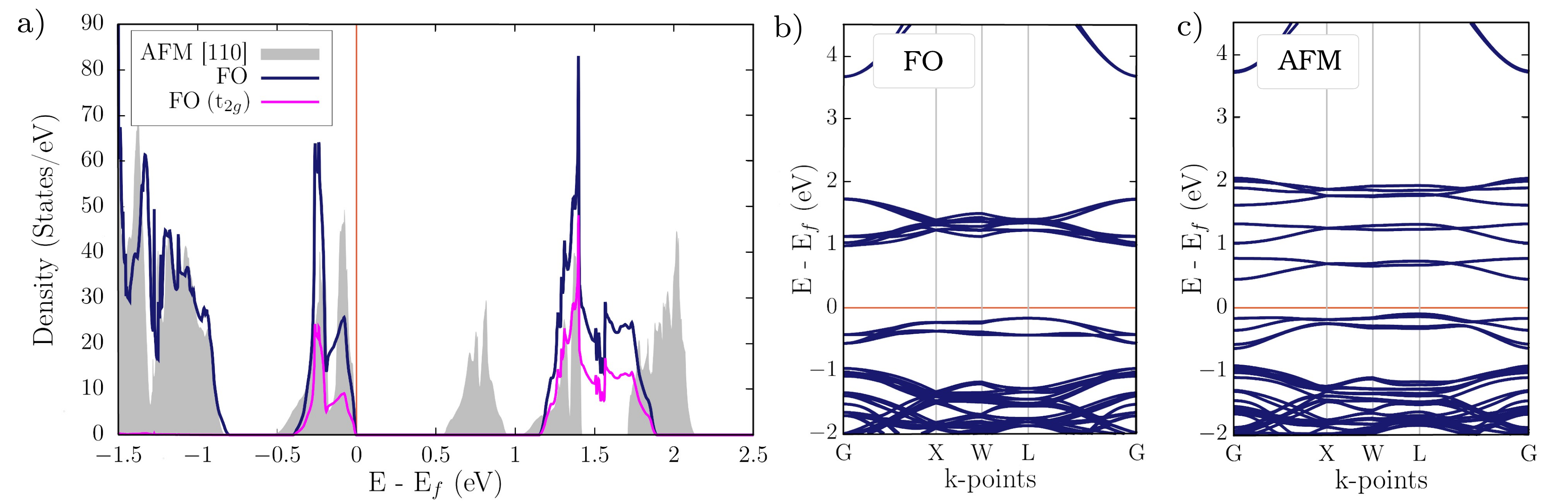}
    \end{center}
\caption{(Color online) Comparison of the electronic properties of BCOO  for the two different magnetic phases FO and AFM-110: (a) density fo states and (b) band structure. 
}
\label{fig:2}
\end{figure*}

After clarifying the basic multipolar character of the FO order with DFT+U+SOC, we move forward to the analysis of the differences between the FO and the competing dipolar AFM-110 solutions in terms of stability, electronic structure and magnetic properties.
Our results show that the FO phase is lower in energy than the dipolar phase, with a difference of $\approx$ 41, 45 and 43 meV/f.u. for BCOO, BMOO and BZOO respectively. 

The density of states and the band structure displayed in in Fig.~\ref{fig:2}
show an insulating electronic ground state for both phases with strong qualitative differences in the character of the unoccupied states:
FO-ordered BCOO exhibits one broad unoccupied t$_{2g}$ band, which is split into three peaks in the AFM phase. The FO insulating energy gap, 1.16 eV, is significantly larger than the corresponding AFM, 0.55 eV, see Table ~\ref{Tab:2}). These are well-defined electronic characteristics that identify a clear difference between these two magnetic orders, which can be verified experimentally.

From a magnetic point of view, the difference between the FO and AFM phases become transparent from the ordered magnetic dipolar moments and the magnetization density. The dipolar AFM-110 posses a local spin and orbital moment on the osmium atom of $ m_{S}\sim$~1.4~$\mu_{B}$, $m_{L} \sim$~0.7~$\mu_{B}$ summing up to $m_{J}\sim$~0.7~$\mu_{B}$, while they average to zero in the FO case. 
The diversity between the FO and AFM-110 magnetic orderings is also reflected in the  magnetization density along z Figure~\ref{fig:1}.

\begin{table}[h!]
	\begin{center}
		\begin{ruledtabular}
			\renewcommand{\arraystretch}{1.2}
			\begin{tabular}{l | c c  }
			    & \multicolumn{2}{c}{Energy Gap (eV)} \\
				 \hline
				Compound & FO & AFM-110 \\   
				\hline
				Ba$_{2}$CaOsO$_{6}$ & 1.16  & 0.55  \\
				
				Ba$_{2}$MgOsO$_{6}$ & 1.16  & 0.52 \\
				
				Ba$_{2}$ZnOsO$_{6}$ & 1.24  & 0.64  \\				
			\end{tabular}
		\end{ruledtabular}
		\caption{\label{Tab:2}  Energy Gap for the different magnetic phases show how the magnetic and electronic properties are strictly related in these compounds. An enhancement of the band gap is found in the multipolar ordered phases, as compared with the conventional dipolar solution.}
	\end{center}
\end{table}

\subsection{Intersite exchanges}

The driving force that couples rank 3 time-odd multipolar operators and stabilizes the multipolar magnetic order in these compounds is the intersite exchange interaction between osmium atoms mediated via superexchange mechanism ~\cite{pourovskii2021}.
To further analyze and compare our result to previous theoretical findings, we start by mapping total energy differences obtained with our constrained ODM approach for distinct multipolar ordered phases, to the following model Hamiltonian
\begin{equation}
    H =\sum_{\langle ij\rangle}\sum_{KQK'Q'}V_{KK'}^{QQ'}(\vR_{ij})O_K^Q(\vR_i)O_{K'}^{Q'}(\vR_j),
\end{equation}
where  $O_K^Q(\vR_i)$ are the Hermitian spherical  tensor~\cite{RevModPhys.81.807} for $J$=2  of the rank $K=1...4$, $Q=-K, ... ,K$ and the sum runs over all nearest-neighbor $\langle ij\rangle$ Os-Os bonds. 
This Hamiltonian is a reduced form of the one used in Ref.~\cite{pourovskii2021}, as the remnant crystal field is already taken into account self-consistently. 
In ref ~\cite{pourovskii2021} the Hamiltonian is further simplified to an effective pseudo-spin Hamiltonian acting on the low lying doublet. Here we follow the same reasoning and rewrite it as 
\begin{equation}
    H =\sum_{\langle ij\rangle}\sum_{\alpha \beta} J_{\alpha \beta} (\vR_{ij}) \tau_{\alpha}(\vR_i) \tau_{\beta} (\vR_j),
    \label{eq:5}
\end{equation}
where $\tau_{\alpha}$ is the corresponding pseudo-spin-1/2 operator and, for $\alpha =$ y, it coincides with O$_{3}^{-2}$.
To calculate the value of the intersite exchange interaction J$_{yy}$ between time-odd multipoles, we employed the constrained ODM for an antiferro-octupolar (AFO) configuration, with the $O_{3}^{-2}$ aligned ferromagnetically in [001] planes and AFM out-of-plane. 
Our result shows that the FO phase is still lowest in energy and that, while the AFO has different signs of the $O_{3}^{-2}$ operators, all other saturated magnetic multipolar moments keep their values unchanged, with a difference of $\sim$ 12 \% in the tensor fit coefficients for $O_{2}^{0}$ and $\sim$ 1 \% for $O_{2}^{0}$.
By assuming that these changes are not as significant as the ones brought by $O_{3}^{-2}$ in the total energy, we calculate the exchange constant as  
\begin{equation}
    J_{yy} =  \frac{E_{FO} - E_{AFO}}{2} .
\end{equation}

Our results, in Table ~\ref{Tab:3}, are in qualitative agreement with FT-HI results but are overestimated by a multiplicative factor ($\approx$ 4), to be attributed to
the different electron counting in the two approaches, as discussed above and in a previous work~\cite{Daje}. Still, the relative strength of the computed $J_{yy}$ in the three compounds is reproduced rather consistently.

\begin{table}[!h]
	\begin{center}
		\begin{ruledtabular}
			\renewcommand{\arraystretch}{1.2}
			\begin{tabular}{l | c c }
				& \multicolumn{2}{c}{J$_{yy}$ (meV/f.u.)} \\ \hline 
				Compound & DFT & FT-HI ~\cite{pourovskii2021} \\   
				\hline
				Ba$_{2}$CaOsO$_{6}$ & -10.70 & -2.98 \\
				
				Ba$_{2}$MgOsO$_{6}$ & -10.10 & -2.93\\
				
				Ba$_{2}$ZnOsO$_{6}$ & -8.47 & -1.71 \\				
			\end{tabular}
		\end{ruledtabular}
		\caption{\label{Tab:3} Comparison between DFT and FT-HI intersite exchange constants for the effective Hamiltonian of equation~\ref{eq:5}.
		}
	\end{center}
\end{table}

\subsection{Tetragonal Distortions}
\setcounter{secnumdepth}{1}
\label{ssec:tet}

We conclude by reporting data on the possibility for these systems to undergo a cubic-to-tetragonal distortion. Osmate DPs have been theoretically predicted to host quadrupolar moments and recently it has been proved that unaxial strain along the z axis might lead to a suppression of the FO transition temperature as a consequence of the interplay of a weakened octupolar exchange interactions and the produced transverse field ~\cite{PhysRevB.104.174431}. 
{Minuscule tetragonal distortions are expected to activate transitions between the ground state and the excited singlet, ultimately detectable in Inelastic Neutron Spectra measurements ~\cite{pourovskii2021}}. We investigated whether such distortions might induce a transition towards a competing favourable solutions by studying the change in energy as a function of the $\delta = c/a -1$ by switching the lattice parameter in steps of 0.01~\AA $ $ in both positive and negative directions from the experimental value, while keeping the volume fixed 
(we further compared our result with an automated conjugate gradient minimization algorithm with both fixed and variable lattice parameters). 

\begin{figure}[t]
  	\begin{centering}
  		\includegraphics[width=1\columnwidth]{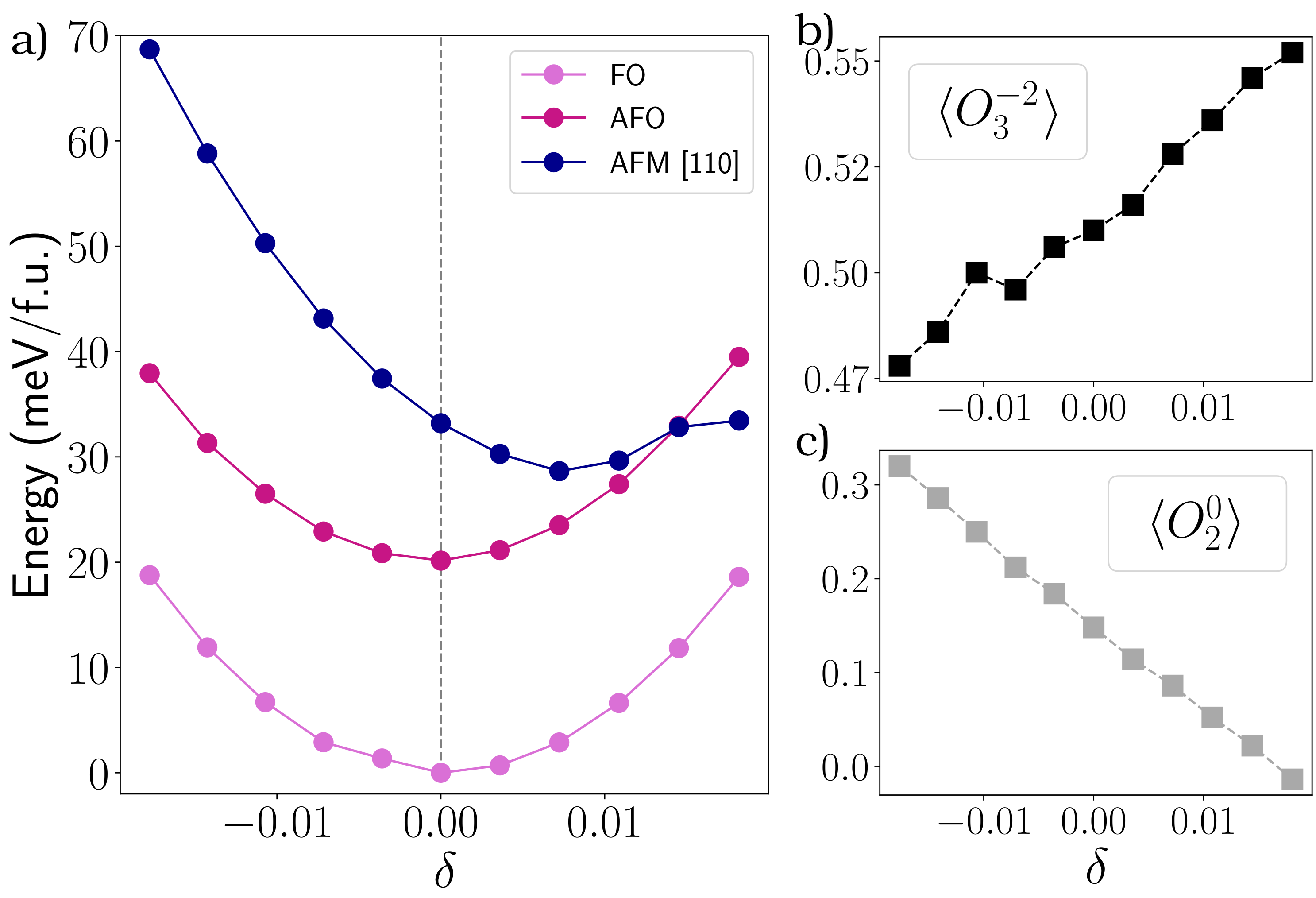}
  		\par\end{centering}
  	\caption{(Color Online) a) Energy as a function of $\delta = c/a -1$ for BCOO. The total energies are scaled to have the zero for the lowest energy value. b) Mean values of the $O_{3}^{-2}$ moment as function of $\delta$ for BCOO. c) Mean values of the $O_{2}^{0}$ moment as function of $\delta$ for BCOO.}
  	\label{fig:3} 
  \end{figure}

The results for BCOO are collected in Figure ~\ref{fig:3} (data for BMOO and BZOO are available in the SM~\cite{supplmat}). 
Our DFT data clearly highlights no deviation from the cubic symmetry for both FM and AFM octupolar  FO and AFO phases.  
In contrast, the AFM-110 phase shows an expansion along the z axis with change in lattice parameter of the order of $\sim $ 1 \% and the appearance of JT distortions, both of which would have been detected by X-ray diffraction measurement. 
These considerations apply also for BMOO and BZOO. 
{An important point is the dependence of the magnetic multipoles on the magnitude of the strain, here analysed for BCOO. We obtain for the $O_2^0$ moment a sharp linear dependence as function of $\delta$ in good qualitative agreement with the DFT+HI results (see Supplementary related to Ref.~\cite{pourovskii2021}).  We further observe a linear behavior of the $O_{3}^{-2}$ multipole which suggests a strengthening of the FO phase upon tensile stress, with a possible enhancement of the corresponding transition temperature (indeed the evaluation of $J_{yy}$ for the $\delta = +0.01$ structure gives a value $\sim$~3~\% larger). For completeness, we find that the multipolar moment $O_2^2$ remains constant.}

These results suggest that tetragonal distortions might play a decisive role on the magnetic properties of these compounds and that  further investigations, also via experimental analysis, might be a  worthy research path.

\section{Conclusion}

In conclusion,  we have proposed a protocol for obtaining magnetic multipolar ordered phases in DFT+U+SOC   based on the constrain of the occupation density matrix as obtained from DFT+DMFT within the FT-HI approximation. We applied this procedure to 5d$^2$ double perovskites, correctly reproducing the FO ordering of time-odd octupoles. We proved the FO phase to be the ground state and compared our results with dipolar AFM-110 configuration,  that would be the "conventional" DFT ground state solution. We calculated the interstite exchange interaction constant, finding reasonable agreement with other theoretical results. Finally, we explored the possibility for these systems to undergo a tetragonal distortion, finding no such evidence 
{and revealed the dependence of the multipolar moments upon strain from purely ab initio perspective}.

\section{Acknowledgments}

D. Fiore Mosca acknowledges the Institut Français d'Autriche and the French Ministry for Europe and Foreign Affairs for the French Government Scholarship as well as the Vienna Doctoral School of Physics.

\bibliography{biblio}

\end{document}


\title{Supplementary material for 
'Modelling Magnetic Multipolar Phases in Density Functional Theory'}

\author{Dario Fiore Mosca}
\affiliation{University of Vienna, Faculty of Physics and Center for Computational Materials Science, Vienna, Austria}

\author{Leonid V. Pourovskii}
\affiliation{Centre de Physique Th\'eorique, Ecole Polytechnique, CNRS, Institut Polytechnique de Paris,
  91128 Palaiseau Cedex, France}
\affiliation{Coll\`ege de France, 11 place Marcelin Berthelot, 75005 Paris, France}

\author{Cesare Franchini}
\affiliation{University of Vienna, Faculty of Physics and Center for Computational Materials Science, Vienna, Austria}
\affiliation{Department of Physics and Astronomy, Alma Mater Studiorum - Universit\`a di Bologna, Bologna, 40127 Italy}

\date{\today}	
\maketitle

\section{First principles methods}

The two-electron Onsite Density Matrix as obtained via FT-HI method (Wien2k) is in spherical harmonic basis $ \big( Y_{2,\uparrow}^{-2} , Y_{2,\uparrow}^{-1},Y_{2,\uparrow}^{0},Y_{2,\uparrow}^{1},Y_{2,\uparrow}^{2},Y_{2,\downarrow}^{-2},Y_{2,\downarrow}^{-1},Y_{2,\downarrow}^{0},Y_{2,\downarrow}^{1},Y_{2,\downarrow}^{2} \big) $ and is given by the following \emph{real part}: 

\vspace{5mm}

\scriptsize
\begin{equation*}
 \begin{pmatrix}
0.2112  & 0  & 0 & 0  & -0.1609 & 0  & -0.1527 & 0 & 0  & 0 \\
0  & 0.4597  & 0  & 0  & 0 & 0 & 0 & -0.0532 & 0 & 0  \\
0 & 0  & 0.0069  & 0  & 0  & 0  & 0 & 0 & -0.0532 & 0 \\
0  & 0  & 0  & 0.1977  & 0 & 0.1093  & 0 & 0 & 0 & -0.1527 \\
-0.1609 & 0 & 0  & 0 & 0.1244  & 0 & 0.1093  & 0  & 0 & 0  \\
0  & 0 & 0  & 0.1093  & 0 & 0.1244  & 0 & 0  & 0 & -0.1609 \\
-0.1527 & 0 & 0 & 0 & 0.1093  & 0 & 0.1977  & 0  & 0  & 0  \\
0 & -0.0532 & 0 & 0 & 0  & 0  & 0  & 0.0069  & 0  & 0 \\
0  & 0 & -0.0532 & 0 & 0 & 0 & 0  & 0  & 0.4598  & 0  \\
0 & 0  & 0 & -0.1527 & 0  & -0.1609 & 0  & 0 & 0  & 0.2113 
\end{pmatrix}
\end{equation*}
\normalsize
\vspace{5mm}

and \emph{imaginary component}

\vspace{5mm}

\scriptsize
\begin{equation*}
 \begin{pmatrix}
0  & 0 & -0.0367 & 0 & 0  & 0  & 0  & 0  & 0.2659  & 0 \\
0  & 0  & 0  & 0.2270  & 0  & 0.1880  & 0 & 0  & 0 & -0.2660 \\
0.0367  & 0 & 0  & 0  & -0.0270 & 0 & -0.0318 & 0  & 0  & 0  \\
0  & -0.2270 & 0 & 0  & 0 & 0 & 0  & 0.0318  & 0  & 0  \\
0 & 0 & 0.0270  & 0  & 0  & 0 & 0 & 0 & -0.1880 & 0  \\
0 & -0.1880 & 0  & 0  & 0  & 0  & 0  & 0.0270  & 0 & 0 \\
0 & 0  & 0.0318  & 0 & 0  & 0 & 0  & 0 & -0.2270 & 0  \\
0 & 0 & 0 & -0.0318 & 0  & -0.0270 & 0  & 0  & 0  & 0.0367  \\
-0.2659 & 0  & 0 & 0 & 0.1880  & 0  & 0.2270  & 0 & 0  & 0 \\
0  & 0.2660  & 0 & 0 & 0 & 0  & 0 & -0.0367 & 0  & 0 
\end{pmatrix}
\end{equation*}
\normalsize

\vspace{0.5cm}

We employed a change of basis from spherical to cubic as in the VASP conventional form  $\big( d_{xy,\uparrow},d_{yz,\uparrow},d_{z^{2},\uparrow},d_{xz,\uparrow},d_{x^{2}-y^{2},\uparrow},d_{xy,\downarrow},d_{yz,\downarrow},d_{z^{2},\downarrow},d_{xz,\downarrow},d_{x^{2}-y^{2},\downarrow} \big)$ using the following transformation matrix (without considering the spin degeneracy)
\vspace{5mm}

\scriptsize
\begin{equation*}
 \begin{pmatrix}
0       & 0.70711 & 0       & 0       & 0 & 0 & 0        & 0       & 0       & -0.70711 \\
0       & 0       & 0       & 0.70711 & 0 & 0 & 0        & 0.70711 & 0       & 0        \\
0       & 0       & 0       & 0       & 1 & 0 & 0        & 0       & 0       & 0        \\
0       & 0       & 0.70711 & 0       & 0 & 0 & -0.70711 & 0       & 0       & 0        \\
0.70711 & 0       & 0       & 0       & 0 & 0 & 0        & 0       & 0.70711 & 0  
\end{pmatrix}
\end{equation*}
\normalsize

\vspace{5mm}

where the columns are alternated real and imaginary parts.
After having added the contribution coming from the e$_{g}$ orbitals (0.61347) and a charge correction on the t$_{2g}$ as explained in the main text we obtain the final occupation matrix given in input with the flag \texttt{OCCEXT} = 1, as reported below.


\vspace{1cm}
\scriptsize
\begin{verbatim}
spin component  1

 0.62239 0.0000 0.04504 0.0000 0.0000       0.0000 0.0000 0.0000 0.0000 0.04340
 0.0000 0.62239 0.0000 0.22700 0.0000       0.0000 0.0000 0.0000 0.13100 0.0000
 0.04504 0.0000 0.60912 0.0000 0.0000       0.0000 0.0000 0.0000 0.0000 0.00686
 0.0000 0.22700 0.0000 0.62239 0.0000       0.0000 -0.13100 0.0000 0.0000 0.0000
 0.0000 0.0000 0.0000 0.0000 0.60912       -0.04340 0.0000 -0.00686 0.0000 0.0000
 
spin component  2

 0.0000 -0.13100 0.0000 0.22695 0.0000       0.0000 0.22695 0.0000 -0.13100 0.0000
 0.13100 0.0000 -0.02249 0.0000 0.03900      0.22700 0.0000 -0.03762 0.0000 -0.02170
 0.0000 -0.02249 0.0000 0.03762 0.0000       0.0000 0.03762 0.0000 -0.02249 0.0000
 0.22700 0.0000 -0.03762 0.0000 0.02170      0.13100 0.0000 -0.02249 0.0000 -0.03900
 0.0000 0.03895 0.0000 -0.02170 0.0000       0.0000 0.02170 0.0000 -0.03895 0.0000

spin component  3

 0.0000 0.13100 0.0000 0.22700 0.0000        0.0000 -0.22700 0.0000 -0.13100 0.0000
 -0.13100 0.0000 -0.02249 0.0000 0.03895    -0.22695 0.0000 -0.03762 0.0000 -0.02170
 0.0000 -0.02249 0.0000 -0.03762 0.0000      0.0000 0.03762 0.0000 0.02249 0.0000
 0.22695 0.0000 0.03762 0.0000 -0.02170      0.13100 0.0000 0.02249 0.0000 0.03895
 0.0000 0.03900 0.0000 0.02170 0.0000        0.0000 0.02170 0.0000 0.03900 0.0000

spin component  4

  0.62244 0.0000 -0.04504 0.0000 0.0000      0.0000 0.0000 0.0000 0.0000 -0.04345
  0.0000 0.62244 0.0000 -0.22700 0.0000      0.0000 - 0.0000 0.0000 -0.13105 0.0000
 -0.04504 0.0000 0.60912 0.0000 0.0000       0.0000 0.0000 0.0000 0.0000 0.00686
  0.0000 -0.22700 0.0000 0.62244 0.0000      0.0000 0.13105 0.0000 0.0000 0.0000
  0.0000  0.0000 0.0000 0.0000  0.60917      0.04345 0.0000 -0.00686 0.0000 0.0000

\end{verbatim}
\normalsize
where the Spin Component 1,2,3,4 refer to ($\uparrow \uparrow$),  ($\uparrow \downarrow$), ($\downarrow \uparrow$) and ($\downarrow \downarrow$) spinor components respectively.
After the final self consistent calculation we obtain the resulting ODMs for BCCO, BMOO and BZOO, which are listed in the following.

\subsection{Final Onsite Density Matrix BCOO}

\normalsize
In the following are listed the final ODM after the self-consistent calculations obtained for BCOO, BMOO and BZOO in the cubic experimental structure.

\scriptsize
\begin{verbatim}
spin component  1

  0.6052 -0.0015 -0.0001 -0.0019  0.0001      0.0000  0.0063 -0.0002 -0.0033  0.0412
 -0.0015  0.6089 -0.0000  0.0022  0.0002     -0.0063 -0.000 -0.0001  0.1293 -0.0002
 -0.0001 -0.0000  0.5944  0.0002  0.0003      0.0002  0.0001 -0.0000 -0.0002 -0.0001
 -0.0019  0.0022  0.0002  0.6159  0.0003      0.0033 -0.1293  0.0002  0.0000 -0.0002
  0.0001  0.0002  0.0003  0.0003  0.5947     -0.0412  0.0002  0.0001  0.0002  0.0000

spin component  2

  0.0005  0.1122 -0.0002 -0.0001  0.0002      0.0003 -0.0002  0.0001 -0.1203  0.0001
 -0.1127 -0.0007 -0.0000  0.0048 -0.0002     -0.0006 -0.0019 -0.0360  0.0033 -0.0205
  0.0002  0.0000  0.0002 -0.0359  0.0001     -0.0002  0.0360 -0.0003  0.0001 -0.0000
 -0.0012 -0.0070  0.0358 -0.0000 -0.0202      0.1188 -0.0038  0.0001  0.0003 -0.0000
 -0.0001 -0.0000  0.0001  0.0204  0.0001     -0.0002  0.0204  0.0001  0.0000 -0.0004

spin component  3

  0.0005 -0.1127  0.0002 -0.0012 -0.0001     -0.0003  0.0006  0.0002 -0.1188  0.0002
  0.1122 -0.0007  0.0000 -0.0070 -0.0000      0.0002  0.0019 -0.0360  0.0038 -0.0204
 -0.0002 -0.0000  0.0002  0.0358  0.0001     -0.0001  0.0360  0.0003 -0.0001 -0.0001
 -0.0001  0.0048 -0.0359 -0.0000  0.0204      0.1203 -0.0033 -0.0001 -0.0003 -0.0000
  0.0002 -0.0002  0.0001 -0.0202  0.0001     -0.0001  0.0205  0.0000  0.0000  0.0004

spin component  4

  0.6064 -0.0014 -0.0002 -0.0037 -0.0001      0.0000 -0.0048  0.0002  0.0027 -0.0412
 -0.0014  0.6109  0.0001 -0.0013 -0.0004      0.0048 -0.0000  0.0000 -0.1291  0.0000
 -0.0002  0.0001  0.5947  0.0001  0.0002     -0.0002 -0.0000 -0.0000 -0.0001  0.0001
 -0.0037 -0.0013  0.0001  0.6138  0.0003     -0.0027  0.1291  0.0001  0.0000  0.0002
 -0.0001 -0.0004  0.0002  0.0003  0.5947      0.0412 -0.0000 -0.0001 -0.0002  0.0000
\end{verbatim}
\normalsize

\clearpage
\subsection{Final Onsite Density Matrix BMOO}

\scriptsize
\begin{verbatim}
spin component 1 
 
 0.6167  0.0001 -0.0134 -0.0001 -0.0000     -0.0000 -0.0001 -0.0000  0.0004  0.0379
  0.0001  0.6165  0.0000 -0.1958 -0.0001      0.0001 -0.0000  0.0000  0.1108  0.0000
 -0.0134  0.0000  0.5843 -0.0000  0.0002      0.0000 -0.0000 -0.0000  0.0000 -0.0013
 -0.0001 -0.1958 -0.0000  0.6170 -0.0000     -0.0004 -0.1108 -0.0000  0.0000  0.0000
 -0.0000 -0.0001  0.0002 -0.0000  0.5844     -0.0379 -0.0000  0.0013 -0.0000  0.0000

spin component  2

 -0.0001  0.1095 -0.0001 -0.1959 -0.0000     -0.0002  0.1972 -0.0000 -0.1104 -0.0000
 -0.1094 -0.0000  0.0066 -0.0001 -0.0116      0.1973 -0.0002 -0.0329 -0.0002 -0.0189
 -0.0001  0.0066 -0.0000 -0.0326  0.0000     -0.0000  0.0328  0.0000 -0.0067 -0.0000
 -0.1959  0.0001  0.0327  0.0000 -0.0188      0.1107 -0.0002 -0.0067 -0.0003 -0.0116
  0.0000 -0.0116  0.0000  0.0189 -0.0000      0.0000  0.0190 -0.0000 -0.0116 -0.0000

spin component  3

 -0.0001 -0.1094 -0.0001 -0.1959  0.0000      0.0002 -0.1973  0.0000 -0.1107 -0.0000
  0.1095 -0.0000  0.0066  0.0001 -0.0116     -0.1972  0.0002 -0.0328  0.0002 -0.0190
 -0.0001  0.0066 -0.0000  0.0327  0.0000      0.0000  0.0329 -0.0000  0.0067  0.0000
 -0.1959 -0.0001 -0.0326  0.0000  0.0189      0.1104  0.0002  0.0067  0.0003  0.0116
 -0.0000 -0.0116  0.0000 -0.0188 -0.0000      0.0000  0.0189  0.0000  0.0116  0.0000

spin component  4

  0.6163 -0.0002  0.0133 -0.0001  0.0001     -0.0000  0.0000  0.0000  0.0003 -0.0379
 -0.0002  0.6163 -0.0000  0.1958  0.0000     -0.0000 -0.0000  0.0000 -0.1105 -0.0000
  0.0133 -0.0000  0.5843  0.0000  0.0002     -0.0000 -0.0000 -0.0000  0.0000 -0.0013
 -0.0001  0.1958  0.0000  0.6170  0.0000     -0.0003  0.1105 -0.0000  0.0000  0.0000
  0.0001  0.0000  0.0002  0.0000  0.5845      0.0379  0.0000  0.0013 -0.0000  0.0000

\end{verbatim}

\subsection{Final Onsite Density Matrix BZOO}

\scriptsize
\begin{verbatim}
spin component  1

  0.6324  0.0007 -0.0155  0.0001 -0.0000      0.0000 -0.0003  0.0000 -0.0001  0.0335
  0.0007  0.6322 -0.0000 -0.2040  0.0000      0.0003 -0.0000 -0.0000  0.1123  0.0000
 -0.0155 -0.0000  0.6512 -0.0000  0.0000     -0.0000  0.0000 -0.0000  0.0000 -0.0016
  0.0001 -0.2040 -0.0000  0.6322 -0.0001      0.0001 -0.1123 -0.0000  0.0000  0.0000
 -0.0000  0.0000  0.0000 -0.0001  0.6512     -0.0335 -0.0000  0.0016 -0.0000  0.0000

spin component  2

  0.0000  0.1123  0.0000 -0.2041  0.0000      0.0000  0.2041  0.0001 -0.1120  0.0000
 -0.1119 -0.0000  0.0077  0.0003 -0.0134      0.2043  0.0000 -0.0290  0.0009 -0.0167
 -0.0000  0.0077  0.0000 -0.0290  0.0000      0.0000  0.0290 -0.0000 -0.0077 -0.0000
 -0.2040 -0.0003  0.0290 -0.0001 -0.0167      0.1124  0.0008 -0.0078  0.0001 -0.0134
 -0.0001 -0.0134  0.0000  0.0167 -0.0000      0.0000  0.0168  0.0000 -0.0135 -0.0000

spin component  3

  0.0000 -0.1119 -0.0000 -0.2040 -0.0001     -0.0000 -0.2043 -0.0000 -0.1124 -0.0000
  0.1123 -0.0000  0.0077 -0.0003 -0.0134     -0.2041 -0.0000 -0.0290 -0.0008 -0.0168
  0.0000  0.0077  0.0000  0.0290  0.0000     -0.0001  0.0290  0.0000  0.0078 -0.0000
 -0.2041  0.0003 -0.0290 -0.0001  0.0167      0.1120 -0.0009  0.0077 -0.0001  0.0135
  0.0000 -0.0134  0.0000 -0.0167 -0.0000     -0.0000  0.0167  0.0000  0.0134  0.0000

spin component  4

  0.6321 -0.0007  0.0155  0.0000  0.0000     -0.0000  0.0003 -0.0000 -0.0001 -0.0335
 -0.0007  0.6321  0.0000  0.2040  0.0000     -0.0003 -0.0000 -0.0000 -0.1123  0.0000
  0.0155  0.0000  0.6512  0.0001 -0.0000      0.0000  0.0000 -0.0000 -0.0000 -0.0016
  0.0000  0.2040  0.0001  0.6322  0.0001      0.0001  0.1123  0.0000  0.0000 -0.0000
  0.0000  0.0000 -0.0000  0.0001  0.6512      0.0335 -0.0000  0.0016  0.0000  0.0000
\end{verbatim}

\clearpage
\section{DOS and Bandstructures}

\subsection{BMOO}

  \begin{figure}[!h]
  	\begin{centering}
  		\includegraphics[width=0.9\columnwidth]{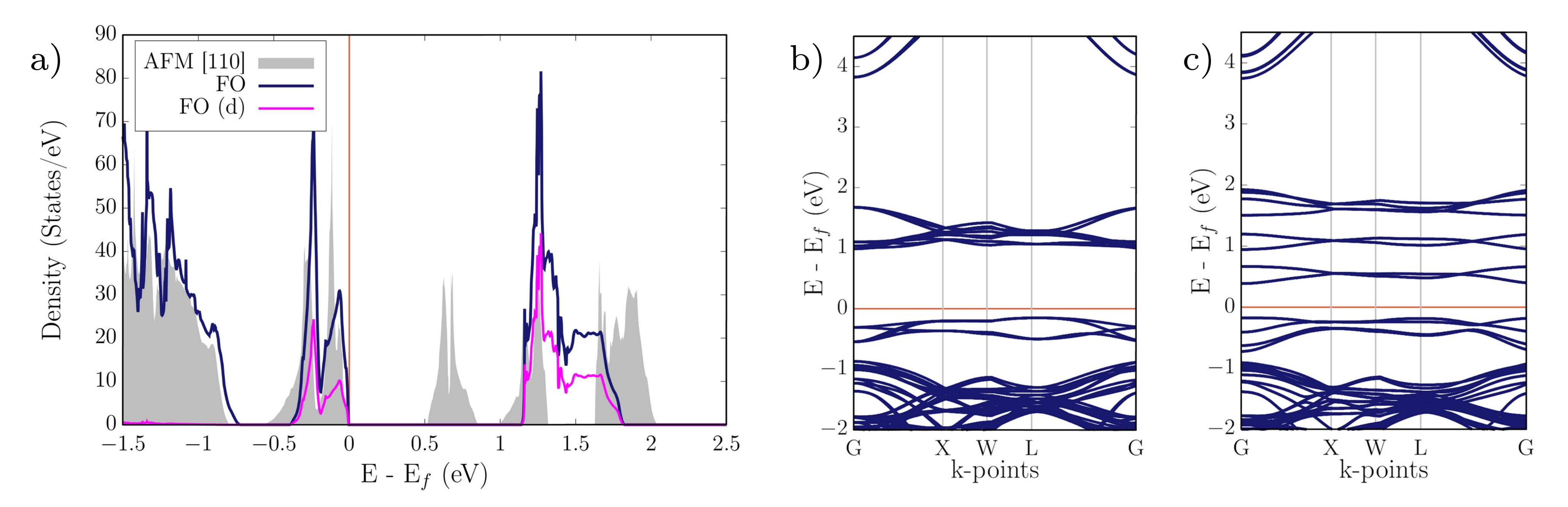}
  		\par\end{centering}
  	\caption{(Color online) Comparison of the electronic properties of BMOO for the two different magnetic phases FO and AFM [110]. The change in Energy Gap is clear in the Density of States (a) and more explicit in the Bandstructures for the FO phase (b) and the AFM [110] phase (c).}
  	\label{fig:3} 
  \end{figure}

\subsection{BZOO}

  \begin{figure}[!h]
  	\begin{centering}
  		\includegraphics[width=0.9\columnwidth]{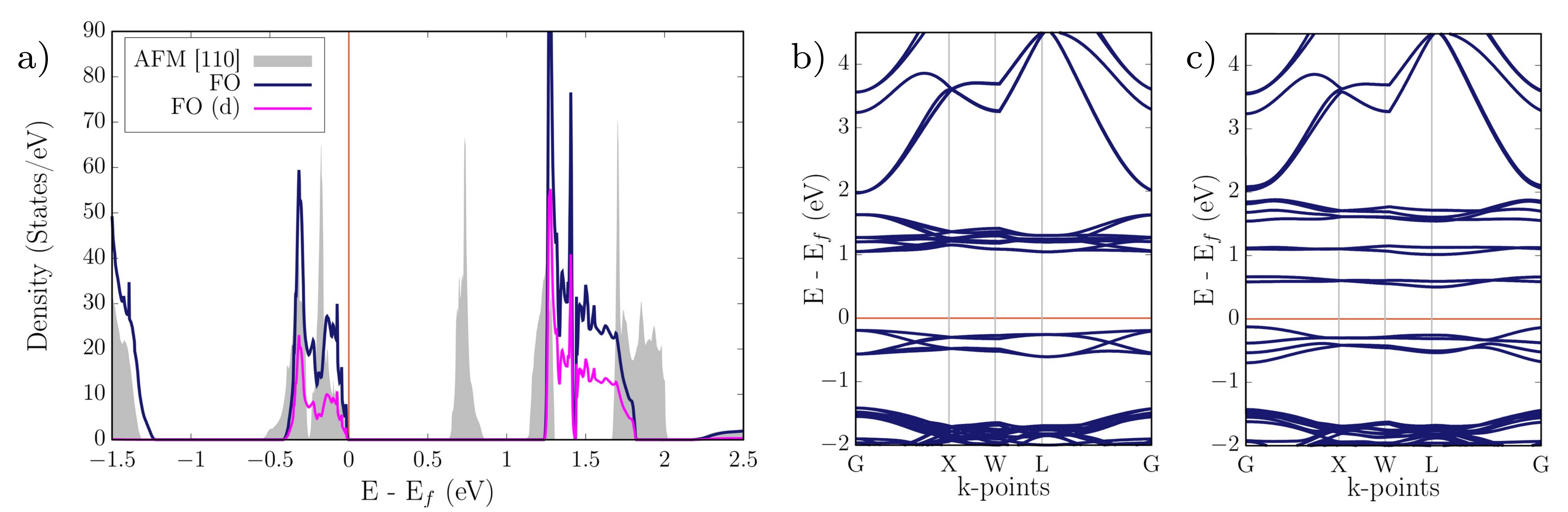}
  		\par\end{centering}
  \caption{(Color online) Comparison of the electronic properties of BZOO for the two different magnetic phases FO and AFM [110]. The change in Energy Gap is clear in the Density of States (a) and more explicit in the Bandstructures for the FO phase (b) and the AFM [110] phase (c). }
  	\label{fig:3} 
  \end{figure}

\clearpage
\section{Magnetization Density Plots x and y components BCOO}

  \begin{figure}[!h]
  	\begin{centering}
  		\includegraphics[width=0.7\columnwidth]{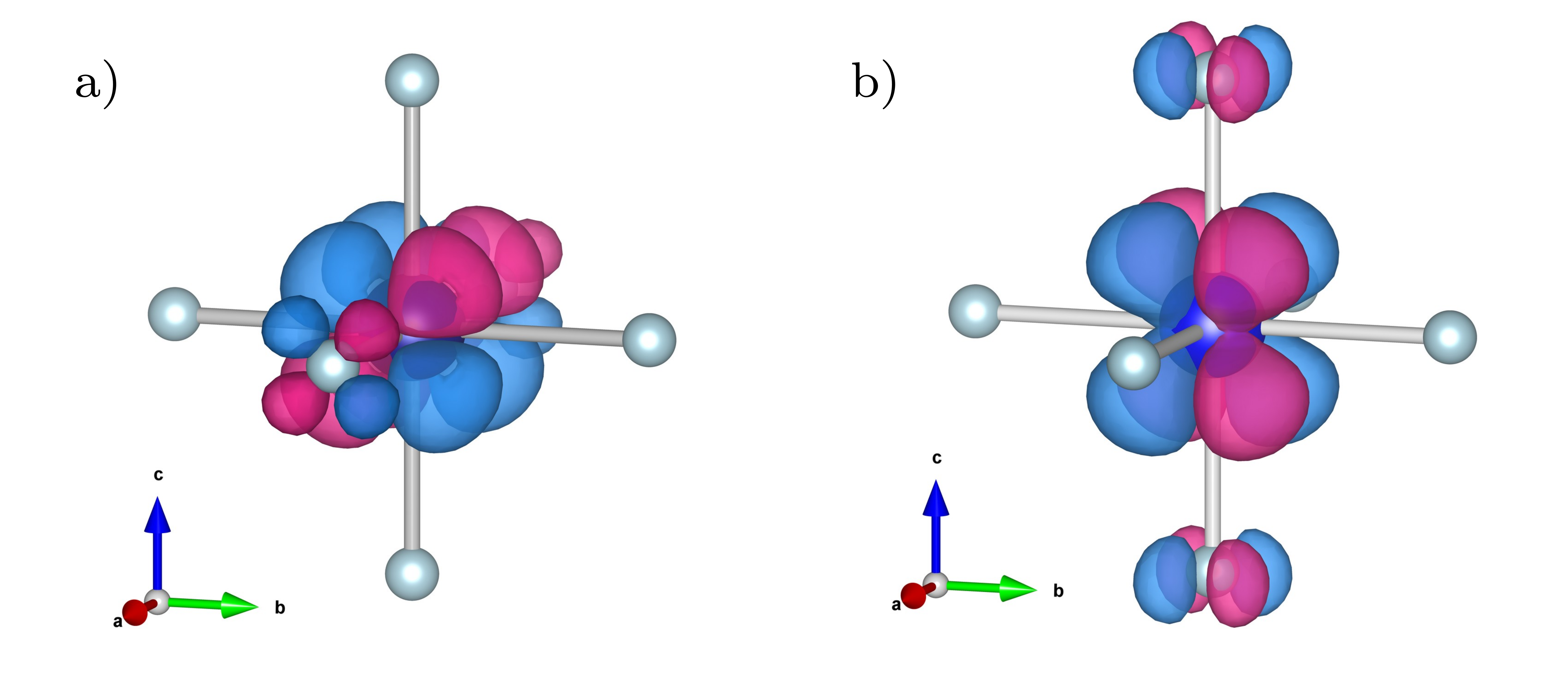}
  		\par\end{centering}
  \caption{(Color online) Plot Magnetization density for the Octupolar FM phase of BCOO of the x and y components in a) and b) respectively.}
  	\label{fig:3} 
  \end{figure}

\section{Tetragonal Distortion BZOO and BMOO}

\subsection{Plot BMOO}

  \begin{figure}[!h]
  	\begin{centering}
  		\includegraphics[width=0.6\columnwidth]{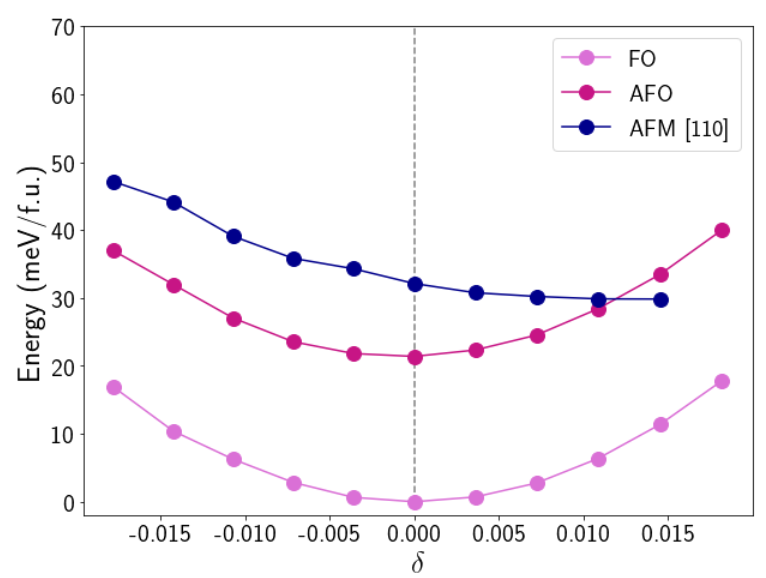}
  		\par\end{centering}
  \caption{(Color online) Energy as a function of $\delta = c/a - 1$ for BMOO. The total energies are scaled to have the zero for the lowest energy value.}
  	\label{fig:3} 
  \end{figure}
\clearpage
\subsection{Plot BZOO}

  \begin{figure}[!h]
  	\begin{centering}
  		\includegraphics[width=0.6\columnwidth]{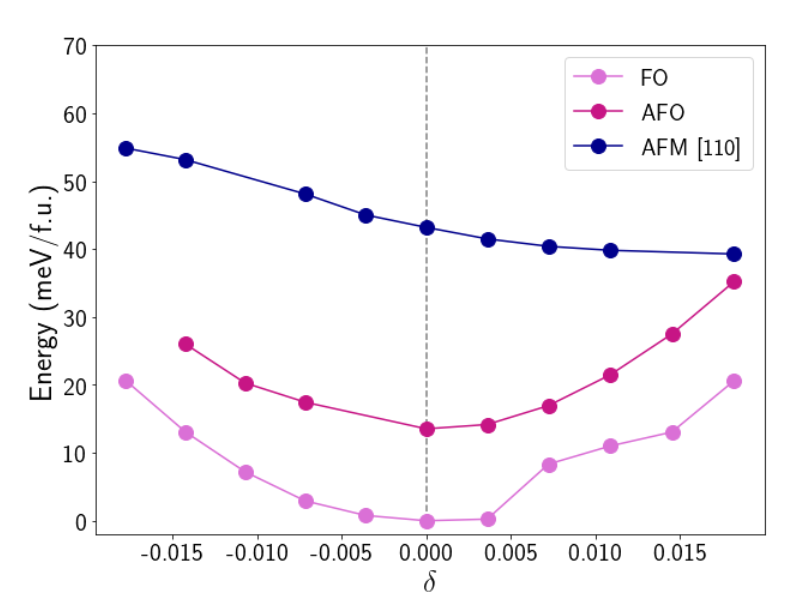}
  		\par\end{centering}
  \caption{(Color online) Energy as a function of $\delta = c/a - 1$ for BZOO. The total energies are scaled to have the zero for the lowest energy value.}
  	\label{fig:3} 
  \end{figure}

\section{Dipolar calculations}

\normalsize 
In the following Table \ref{Tab:magconf} are listed the total energies for  dipolar configurations calculated within DFT+U+SOC using the +U correction in the rotationally invariant Dudarev's scheme and U = 3.4 eV.

\noindent The "xy" notation refers to magnetic moments aligned ferromagnetically within the xy planes along the [abc] crystallographic direction, and anti-ferromagnetically in between adjacent xy planes. Otherwise the magnetic moments are to be read as anti-ferromagnetically aligned within the xy plane, with magnetic moments pointing along the [abc] crystallographic direction, and ferromagnetically aligned in adjacent xy planes.

\begin{table}[!h]
	\begin{center}
			\renewcommand{\arraystretch}{1.2}
			\begin{tabular}{l | c }
                Configuration & Energy (eV) \\
                \hline
                \hline
                AFM [001]          & -66.683  \\
                AFM-xy [001] & non converged   \\
                AFM [100]          & -66.710    \\
                AFM-xy [110] & -66.710    \\
                AFM [110]          & -66.708  \\
                AFM-xy [110] & -66.713  \\
                AFM [111]          & -66.688    \\
                AFM-xy [111] & -66.708    \\
                FM [001]           & non converged    \\
                FM [100]           & -66.696    \\
                FM [110]           & -66.701    \\
                FM [111]           & -66.701   
			\end{tabular}
		\caption{\label{Tab:magconf} Comparison total energies calculations with dipolar magnetic configuration.}
	\end{center}
\end{table}